\documentclass[aps,prd,twocolumn]{revtex4}
\usepackage{amsmath,amssymb,bm}
\usepackage{graphicx}
\usepackage{epstopdf}
\usepackage{amsfonts}
\usepackage{amssymb}
\usepackage{amsbsy}
\usepackage{amsmath}
\usepackage{latexsym}
\usepackage{sansmath}
\usepackage{subfigure}
\usepackage{lipsum}
\usepackage{float}
\usepackage{cancel}

\usepackage{bm}
\usepackage{xcolor}
 
\usepackage{comment}
\usepackage{csquotes}								
\usepackage{physics}
\usepackage{accents}

\def\bea {\begin{eqnarray}}
\def\eea {\end{eqnarray}}
\def\nn {\nonumber}

\begin{document}
\title{Semiclassical Mixmaster Universe}

\author{Muhammad Muzammil} \email{muhammad.muzammil@unb.ca}
\author{Viqar Husain} 
\email{vhusain@unb.ca}
\affiliation{Department of Mathematics and Statistics, University of New Brunswick, Fredericton, Canada.\\}

\begin{abstract}
We present a semiclassical study of the Mixmaster cosmology minimally coupled to a massive scalar field in the Hamiltonian formalism, with focus on three distinct scenarios: the classical cosmology coupled to the quantized scalar field, and ``effective" cosmology, with spacetime discreteness corrections, coupled to the classical scalar field, and to the quantized scalar field. We find several results: (i) the effective cosmology undergoes several small bounces before expanding, with scalar field excitations rising through the bounce; (ii) anisotropies rise and fall as the universe undergoes a bounce, a feature that is enhanced with matter; (iii) Lyapunov exponents reveal that chaos is reduced in the effective systems compared to the classical case.

\end{abstract}

\maketitle  

\section{Introduction} 
The behavior of the universe  close to the big bang singularity has been a question of much interest over the past decades. Despite several notable works, a  widely accepted picture has yet to emerge. Seminal papers by Belinski, Khalitnikov, and Lifshitz, \cite{belinskii1969nature, BKL1971} showed that the approach to a spacelike singularity in Einstein gravity is homogeneous, dominated by time derivatives, and therefore described as a dynamical system  of coupled non-linear ordinary differential equations (ODE). This led to the study of  homogeneous anisotropic Bianchi models that describe the evolution of scale factors moving in potentials that come from spatial curvature \cite{Ryan:1975jw}. The most interesting of these is the Bianchi IX model, also termed the ``Mixmaster Universe" \cite{misner1969mixmaster}. It describes the universe as a billiard ball in a triangular potential with exponentially high expanding wall. Between the walls, the potential is negligible and the universe is described by the Kasner spacetime $ ds^2=-N^2dt^2+\sum_{i=1}^3 t^{2p_i}dx_i^2$,
where the so-called Kasner exponents $p_i$  satisfy the sum rules $\sum_i p_i=1$ and $\sum_i p_i^2=1$ by virtue of the Einstein equations. These exponents transition to new ones at each collision with a wall of the potential, sending the universe into a new Kasner epoch with different exponents. The transitions are described by rules that have been the subject of many studies in vacuum \cite{qcIMisner}  and  with matter coupling \cite{MMdynamicHusainAli}.  This interesting dynamics exhibits chaos, a feature that has led to many investigations over the years \cite{JDBarrow,JDBarrow2, hobill1991mixmaster, burd1990numerical,kiefer2018dynamics, bini2009electrocardiogram, andriopoulos2008mixmaster}.

Despite these studies, the  indicators that quantify chaos fail to be consistent when applied to the Mixmaster universe, giving varying numerical results; see e.g. \cite{nochaos001Ferraz,chaosAdamKrawiec}.  Furthermore, a question of covariance also arises: is chaos an actual physical feature or merely a coordinate artifact? This question has been addressed in by using the Kasner transition rules as an iterative map \cite{CornishLevin, FareyTale}, and by rewriting the evolution equations as a geodesic equation on an auxiliary \cite{MontaniCovariancelyapu}. Both these methods confirm that the Bianchi IX model is chaotic.

The work presented here takes these past analyses further by including the quantum gravity effects of a bounce and quantization of matter, effects that should be important near the cosmological singularity. Several approaches to quantum gravity including supergravity indicate that  chaotic dynamics persists with quantum effects \cite{stringMarkHanneux}. There is a formulation of the BKL conjecture in the connection-triad formalism  \cite{ashtekarhamBKL} using variables introduced in \cite{Barnich:1996sh}, and  studies of  singularity resolution in anisotropic models in symmetry reduced models in the Loop Quantum Gravity (LQG) approach \cite{wilson2018loop, wilson2019quantum, blackmore2023numerics, wilson-ewingLQCbianchiIX} (termed Loop Quantum Cosmology (LQC)). Another recent approach to semiclassical cosmological models is an effective approach where a truncated set of coupled classical equations for expectation values of products of powers of configuration and momentum operators in semiclassical states are studied \cite{bojowaldbrizuelaART, brizuelabojowald, brizuelaUria}--these works indicate that chaos persists but is reduced in this semiclassical approach to the Bianchi IX model.

In this work we present a comparative study of three different approaches to ``effective" and ``semiclassical"  cosmology applied to the Bianchi IX cosmology. The first is a canonical approach to the semiclassical Einstein equation, where gravity is classical and matter is quantum with an evolving state, a system that has full quantum matter backreaction on a classical metric; such an approach has been studied for homogeneous isotropic cosmology \cite{HusainSinghFriedmanSchroedinger, HusainSinghQBRCU}; the second is the addition of spacetime discreteness using the LQG-inspired polymer quantization of volume, both with and without the quantization of a scalar field. In all cases we compare the resulting dynamics with the corresponding classical Bianchi IX model.

The structure of this paper is as follows: In section II, we briefly review the canonical Mixmaster model, introduce the volume discreteness correction, and the effective Hamiltonian with quantum scalar field. In section III, we present numerical results  for the three cases under study, together with results on characterizing  chaos. We conclude in the final section with a summary and discussion of our results.

\section{Mixmaster Cosmology}

The starting point of our analysis is the homogeneous anisotropic line element  
\bea
ds^2=-N^2(t)dt^2 + \sum_{k=1}^3 a_k^2(t)dx^k.
\eea
Misner's parameterization \cite{misner1969mixmaster,Ryan:1975jw} replaces the three scale factors by the volume and anisotropy variables  $(\Omega,\, \beta_+, \, \beta_-)$ defined by 
\bea 
\nu &=& \ln(a_1a_2a_3) \label{omega-vol},\\
\beta_+ &=& \frac{1}{3} \ln \left(\frac{a_1a_2}{a_3^2}\right) \label{bp},\\
\beta_- &=& \frac{1}{\sqrt{3}}\ln \left(\frac{a_1}{a_2}\right)\label{bm},
\eea
The canonical action for the Bianchi IX model with scalar field in these variables is 
\bea 
S &=& \int dt\ \biggr[p_\nu\dot{\nu}+  p_+\dot\beta_+ + p_-\dot\beta_- + p_\phi \dot \phi\nn\\
&& +N(H_{IX}+H_\phi)\biggr];
\eea
where ($p_\pm,p_\nu$) are the momenta conjugate to  ($\beta_\pm,\nu$), $(\phi,p_\phi)$ are the scalar field canonical variables, $N$ is the lapse function, and 
\bea
\label{SFclass}H_\phi &=&  \frac{p_\phi^2}{2\nu}+\nu \chi(\phi) , \\
H_{IX} &=& -\frac{3}{8}\nu p_\nu ^2 + \frac{1}{24\nu}\left(p_+^2+p_-^2\right) \nn\\
&& +\nu^{1/3}\ V_{IX}(\beta_+, \, \beta_-),\label{IXclass}\\
V_{IX}(\beta_+,\, \beta_-)&=& e^{-8\beta_+}+2e^{4\beta_+}\left[\cosh(4\sqrt{3}\beta_-)-1\right] \nn\\
&&-4e^{-2\beta_+}\cosh(2\sqrt{3}\beta_-). \label{Mix-pot}
\eea 
We restrict to the massive scalar field $\chi(\phi)=\frac{1}{2}m^2 \phi^2$. 

The evolution equation equations following from the Hamiltonian constraint $H_C \equiv H_{IX} + H_\phi$ describe the universe as a point moving in a triangular potential $V_{IX}(\beta_+,\beta_-)$, which has exponentially high walls; the volume prefactor in (\ref{Mix-pot}) scales $V_{IX}$, shrinking the triangular potential as the singularity $\nu\rightarrow 0$ is approached. When the universe point is  away from the potential walls, the dynamics is given by the Bianchi-I (Kasner) solution, which corresponds to  $V_{IX}=0$. 

The Kasner Hamiltonian constraint in the phase space variables $(a_k,\pi_k)$ (see e.g. \cite{Ali_2017,muzhusain001}) is 
\bea 
H_I = \frac{1}{4v}\left[ \sum_k \mu_k^2 - \frac{1}{2}\left(\sum_k \mu_k\right)^2\right],
\eea
where $\mu_k=a_kp_k$ and $v= a_1a_2a_3= e^\nu$; the evolution equations are 
\bea  
 \dot{\mu}_i &=& \{ \mu_i,\, H_K\} =0, \label{ldot}\\
 \dot{v} &=& \{v,\, H_I\} = -\frac{1}{4}\sum_k \mu_k \equiv \sigma \label{vdot} \\
h_k &\equiv& \frac{\dot{a}_k}{a_k}  = \frac{1}{2v} \left(\mu_k + 2\sigma\right) \label{hdot}.
\eea
From these it follows that $\mu_k$ are constants of the motion, 
\bea 
v(t) &=& \sigma t  + v_0, \label{nu_kasner}\\
a_k(t) &=& a_k(0) \left[ \nu(t) \right]^{p_k}, \quad p_k \equiv 1+\frac{\mu_k}{2\sigma};\label{Kexp}
\eea 
$\Sigma_{k=1}^3 p_k=1$ follows as an identity from the definition of $p_k$, and the hamiltonian constraint $H_K=0$ gives the non-trivial sum rule $\sum_{k=1}^3\,p_k^2=1$.
 
The exponents $p_k$ thus specify a Kasner solution. A ``Kasner transition" occurs when the universe point bounces off a wall of the potential, changing  the exponents to new values.  The number of Kasner transitions increase as the universe approaches the final singularity.  

\subsection{Effective equations}

The main ingredient for defining effective equations arises from the idea of fundamental spacetime discreteness. In LQG this comes from the kinematical Hilbert space used for quantization, which is based on quantum states defined on graphs. This leads to a discrete spectra of area and volume operators. At the computational level in LQC, discreteness of volume amounts to the introducing of a volume lattice in the classical theory, followed by quantization of the discrete classical theory. An important element of the quantization is that the momentum conjugate to the volume is not defined as a derivative operator, but must be derived from the lattice translation operator. This feature makes the volume momentum bounded above.  

This is formalized by introducing an alternative to Schr\"{o}dinger quantization known as polymer quantization \cite{Halvorson_2004, Ashtekar_2003,Fredenhagen_2006,Velhinho_2007}. The Hilbert space is the span of the basis set $\psi_\nu(x) =\langle x|\nu \rangle= e^{i\nu x}$ with inner-product 
\bea
\langle \nu' |\nu \rangle = \lim_{L\rightarrow \infty} \frac{1}{2L} \int_{-L}^L dx\ e^{-i\nu' x}e^{i\nu x} = \delta_{\nu',\nu}.
\eea
The volume and translation operators  act on this space as
\bea
\hat{v} |\nu\rangle &=& |\nu \rangle \nn\\
\hat{U}_\lambda &\equiv& \widehat{e^{i\lambda v}}|\nu\rangle = |\nu +\lambda\rangle.
\eea
With this inner-product, the state $\psi(x) =\sum_{k=1}^N c_k e^{ix \nu_k}$ is non-zero  at the $N$ points $\nu_k$; a Schr\"{o}dinger state, such as a Gaussian, would require an uncountable superposition in this basis. Therefore in practical calculations, a lattice with fixed spacing is used, corresponding to a fixed $\lambda$ in the translation operation. A ``momentum operator" is constructed from the translation operator via, for instance $\hat{p}_\lambda := (\hat{U}_\lambda - \hat{U}_\lambda^\dagger)/(2i\lambda)$.  

An effective classical theory is obtained by starting with Gaussian states peaked at a particular phase space point, taking expectation values of the Hamiltonian operator defined on the lattice, and imposing a Poisson bracket on the peaking values \cite{Husain:2006uh,Taveras:2008ke,Saeed:2024gtk}.   This procedure  amounts to the replacement  
\bea 
p_\nu \, \to \, \frac{\sin(\lambda p_\nu)}{\lambda}
\label{poly_v}
\eea
in the Mixmaster Hamiltonian constraint (\ref{IXclass}), giving  
\bea 
H_{IX}^{\rm{eff}} &\equiv& -\frac{3}{8\lambda^2}\nu\sin^2(\lambda p_\nu)+\frac{1}{24\nu}\left(p_+^2+p_-^2\right) \nn\\
&& \qquad +\nu^{1/3}\ V_{IX}(\beta_+, \,\beta_-) \label{H-poly-mm}=0.
\eea 
In arriving at this expression, we made the choice that only the volume momentum is bounded above due to polymer quantization, and not the anisotropy variables or scalar field. This is motivated by work in the homogeneous isotropic models where it was found that the volume variable is responsible for the universe undergoing a bounce \cite{agullo2016loopquantumcosmologybrief, bodendorfer2016elementaryintroductionloopquantum, Saeed:2024gtk}. A similar effective approach that includes isotropy variables has been studied \cite{MontaniALLpolymer}.

\subsubsection{Effective cosmology with classical scalar field}
The dynamics of the first model we consider follows from the effective constraint
\bea
H_C^{\rm{eff}} &\equiv& H_{IX}^{\rm{eff}}+H_\phi =0, \label{hamclasspoly}
\eea
where the scalar field part of the Hamiltonian constraint remains unmodified and $H_{IX}^{\rm{eff}}$ is defined in eqn.(\ref{H-poly-mm}). The equations of motions are  
\bea
\dot \nu &=& \{ \nu, \, H_C^{\rm{eff}}\}=-\frac{3}{8}\nu\,\frac{\sin(2\lambda p_\nu)}{\lambda},\label{sc-vdot}\\
\dot p_\nu &=&\{ p_\nu, \, H_C^{\rm{eff}}\}= \frac{3}{8}\frac{\sin^2(\lambda p_\nu)}{\lambda^2}+\frac{1}{24\nu^2}(p_+^2+p_-^2)\nn\\
&&-\frac{1}{3}\nu^{-2/3}V_{IX}(\beta_+,\,\beta_-)\nn\\
&&+\left(\frac{p_\phi^2}{2\nu^2}-\frac{1}{2}m^2\phi^2\right),\\
\dot\beta_\pm &=& \{ \beta_\pm, \, H_C^{\rm{eff}}\}= \frac{1}{12\nu}\,p_\pm\\
\dot p_\pm &=& \{ p_\pm, \, H_C^{\rm{eff}}\}=  \nu^{1/3}\ \frac{\partial}{\partial \beta_\pm}V_{IX}(\beta_+,\,\beta_-)\\
\dot \phi &=& \{ \phi, \, H_C^{\rm{eff}}\}= p_\phi/\nu\\
\dot p_\phi &=& \{ p_\phi, \, H_C^{\rm{eff}}\}= -\nu\chi'(\phi).
\eea 

\subsubsection{Effective cosmology with quantum scalar field}
The second model we study uses an effective constraint where the scalar field is quantized its quantum state evolves via the time dependent Schr\"{o}dinger equation. This may be viewed as the canonical version of the semiclassical Einstein equation applied to cosmology, but with the polymer quantization corrections to the gravitational part. The effective constraint is now
\bea
H_Q^{\rm{eff}} &\equiv& H_{IX}^{\text{eff}}+\langle \psi|\hat{H}_\phi |\psi\rangle =0, \label{hamquantpoly}
\eea
together with  
\bea
i|\dot{\psi}\rangle &=&\hat{H}_\phi|\psi\rangle. \label{t-schro}
\eea
The gravitational Hamilton equations are identical to those $H_C^{\rm{eff}}$ , with the exception of the $\dot{p}_\nu$ equation, which now reads
\bea  
\dot p_\nu &=& \{ p_\nu, \, H_Q^{\rm{eff}}\}= \frac{3}{8}\frac{\sin^2(\lambda p_\nu)}{\lambda^2}+\frac{1}{24\nu^2}\ (p_+^2+p_-^2)\nn\\
&&-\frac{1}{3}\nu^{-2/3}\ V_{IX}(\beta_+,\,\beta_-)-\frac{\partial}{\partial \nu}\langle \psi|\hat{H}_\phi |\psi\rangle\nn \label{pv-dot-qc-IX} 
\eea
The last term  of this equation can be expanded to give
\bea 
\frac{\partial}{\partial \nu}\langle \psi|\hat{H}_\phi |\psi\rangle &=& \frac{1}{\dot \nu}\left(\langle \dot \psi|\hat{H}_\phi |\psi\rangle + \langle \psi|\hat{H}_\phi \dot\psi\rangle\right) \nn \\
&& +\langle \psi|\frac{\partial}{\partial \nu}\hat{H}_\phi |\psi\rangle \nn \\
&=& -\frac{1}{2}\left(\frac{\langle \hat p_\phi^2\rangle}{\nu^2}-m^2 \langle \hat \phi^2 \rangle\right). 
\eea  
It can be verified that the hamiltonian constraint $H_Q^{\rm eff}$ is preserved: labeling the set of the configuration variables $\alpha_I=\{\nu, \beta_\pm\}$ and their canonically conjugate momenta $\pi_J=\{p_\nu,\, p_\pm\}$,   
\bea 
\frac{d}{dt}H_Q^{\rm{eff}} &=& \sum_I\frac{\partial H_Q^{\rm{eff}} }{\partial \alpha_I }\dot\alpha_I+\sum_J\frac{\partial H_Q^{\rm{eff}}}{\partial \pi_J }\dot\pi_J\nn\\
&+&\langle \dot{\psi}|\hat{H}_\phi|\psi \rangle+\langle \psi |\hat{H}_\phi|\dot{\psi}\rangle \nn\\
&=&0.
\eea 
It is similarly verified that probability is conserved ensuring that the complete set of equations is self-consistent. 

In the following we study numerically the dynamics of six models: the zero scalar field classical and effective cases $H_{IX}$ and $H_{IX}^{\rm{eff}}$; the non-zero classical and quantum scalar field cases $H_C = H_{IX} + H_\phi$ and $H_Q = H_{IX} + \langle \psi| \hat{H}_\phi|\psi\rangle$; and $H_C^{\rm{eff}}$ and $H_Q^{\rm{eff}}$, where the volume momentum term comes from polymer quantization. The cosmological bounce is due to this last feature as may be seen by deriving the effective Hubble rate $H_\nu = \dot{\nu}/\nu$ from  (\ref{sc-vdot}) and any one of effective Hamiltonian constraints:
\bea
H_\nu^2 = \frac{3}{2}\ \rho_T\left(1-\frac{\rho_T}{\rho_{\rm{crit}}} \right)
\eea 
where $\rho_{\rm{crit}} = 3/8\lambda^2$ and $\rho_T$ is the sum of the energy densities of the scalar field and anisotropies, e.g.   
\bea 
\rho_T = \frac{1}{\nu}\left( \frac{1}{24\nu}(p_+^2+p_-^2) + \nu^{1/3}\ V_{IX} + \frac{1}{\nu} H_\phi\right)
\eea
for the $H_C^{\rm{eff}}$ model. 

\section{Numerical Results}

\begin{figure*}[hbtp]
    \centering
   \includegraphics[width=1\textwidth]{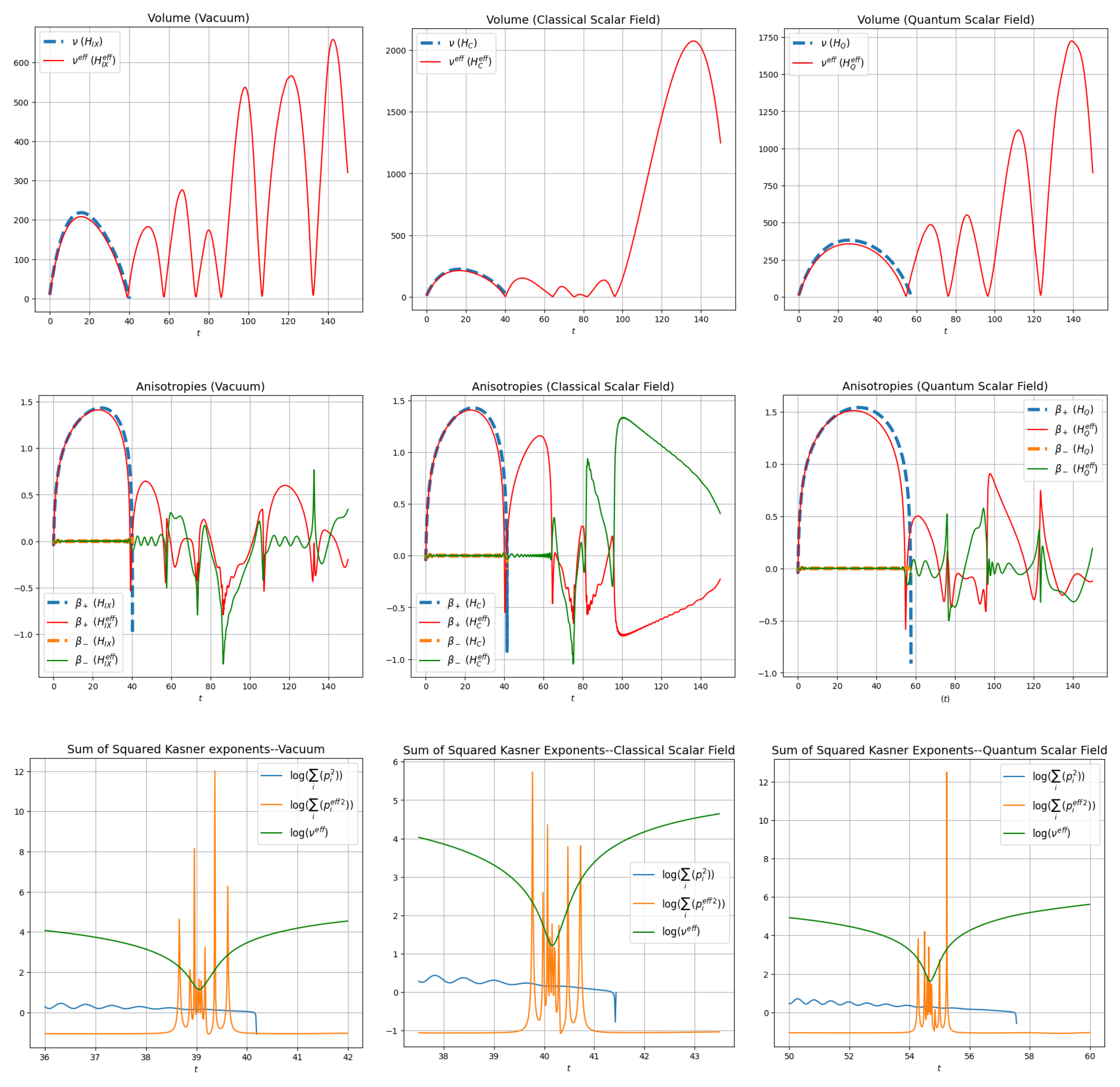}
    \caption{A typical evolution for the classical and effective Mixmaster universe in vacuum, and with classical and quantum scalar field. The dashed blue line is for the classical Bianchi IX Hamiltonian; the solid line  represents the effective Bianchi IX Hamiltonian. The first row shows multiple bounces in the volume for the effective universe. The second row shows the evolution of the anisotropies $\beta_\pm$, and the third, the log of the Kasner squared sum rule overlayed with log of the volume.}
    \label{fig01:dynamics}
\end{figure*}

\begin{figure*}[hbtp]
    \centering
   \includegraphics[width=1\textwidth]{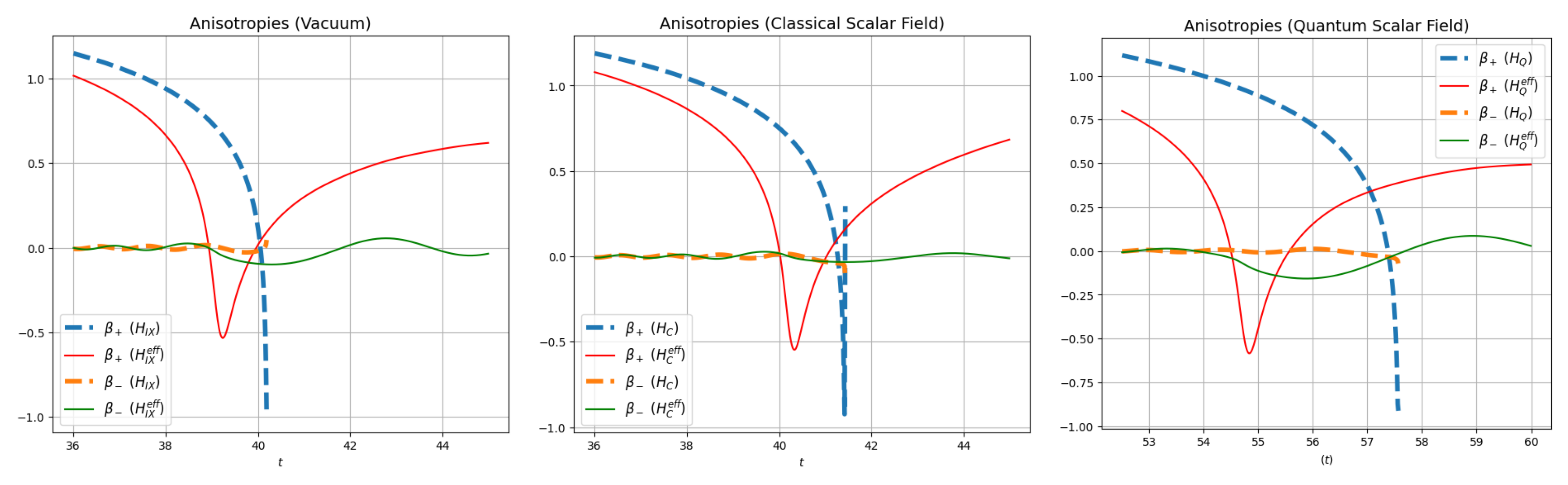}
    \caption{Anisotropy variables $\beta_\pm$ near the first bounce point for each of the models studied.}
    \label{fig09:zoomed_in_anisotropies}
\end{figure*}

\begin{figure*}[hbtp]
    \centering
   \includegraphics[width=0.8\textwidth]{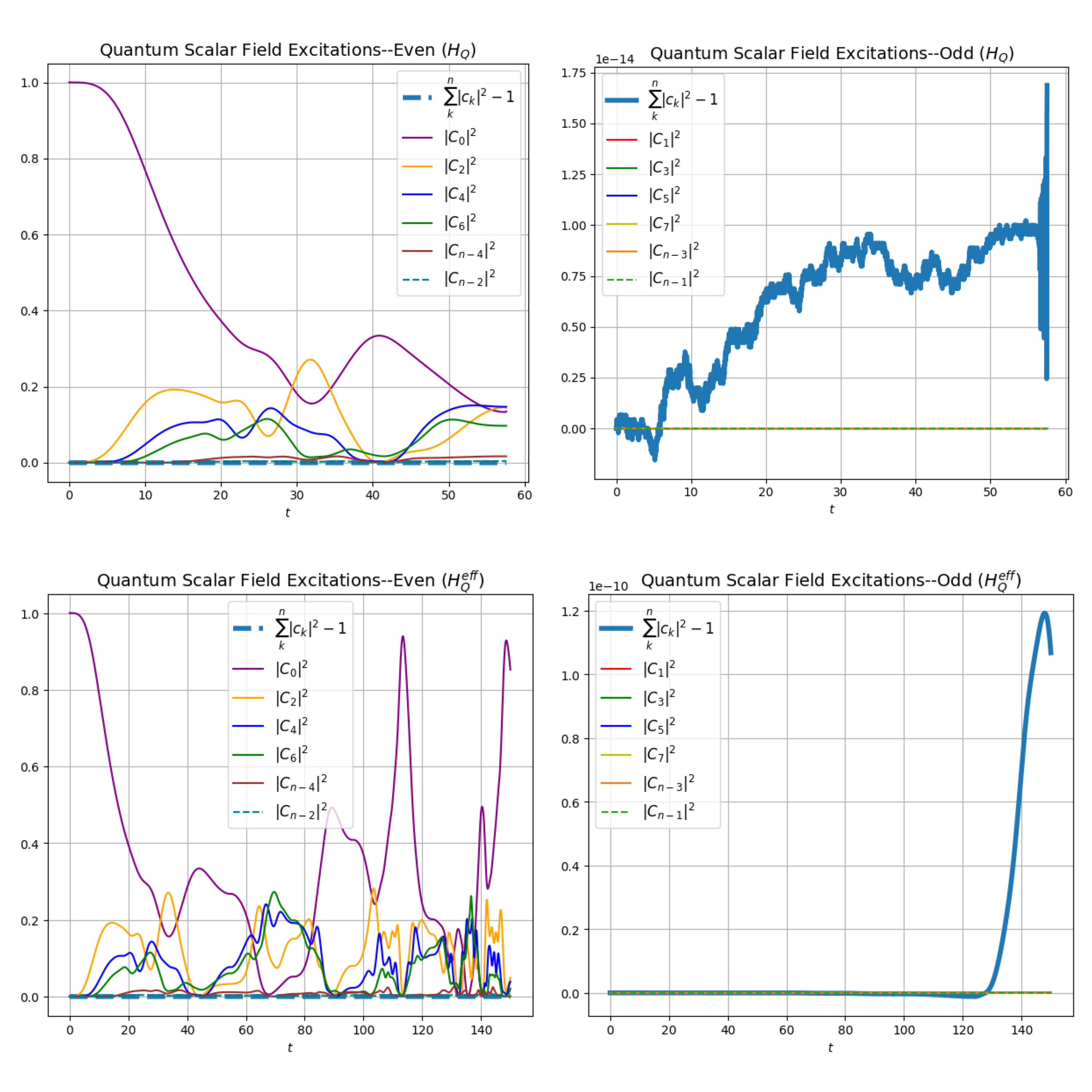}
        \caption{Scalar field state excitations for the Hamiltonians $H_Q$ (classical gravity) top row, and $H_Q^{\rm{eff}}$ (effective gravity) bottom row, starting from the initial ground state. Both show excitations from the ground state as the singularity/bounce is approached,  de-excitation as the universe expands after the bounce, and probability conservation.}
    \label{fig02:excitations}
\end{figure*}

\begin{figure*}[hbtp]
    \centering
   \includegraphics[width=0.85\textwidth]{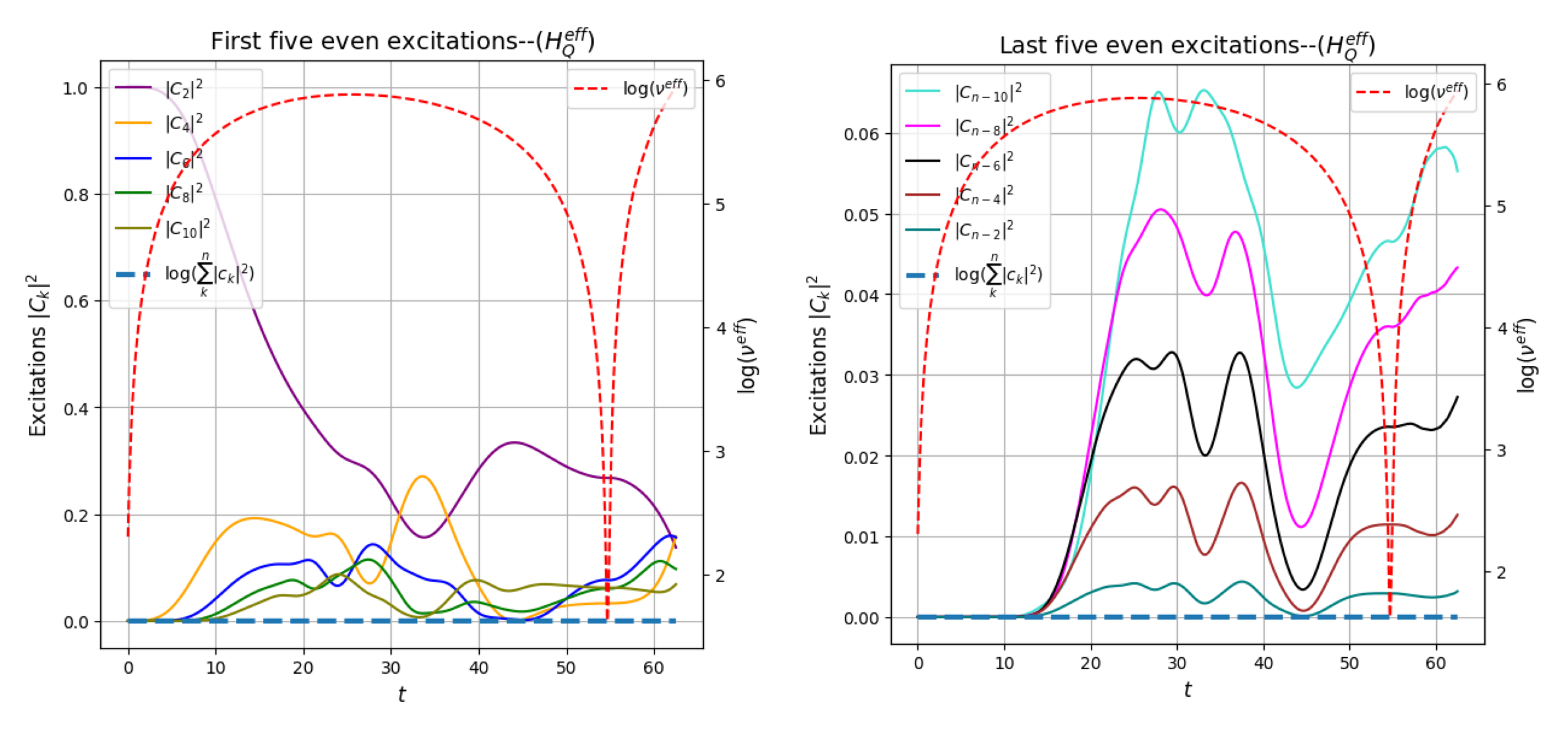}
    \caption{The first few and and the last few even-state excitations of the scalar field (with Hilbert space truncation $n=30$) through the first bounce of the effective universe; the volume is  overlaid (in red).  }
    \label{fig03:exciations_detail}
\end{figure*}

We used the Runge-Kutta-4 method to solve the set of coupled evolution equations for the six models. For useful comparisons of evolution, it is important to set up initial data that is as close as possible. This is straightforward for the vacuum and classical scalar field models $H_{IX}, H^{\rm{eff}}_{IX}$ and $H_C,H^{\rm{eff}}_C$ respectively, since for these cases the initial data is solely configuration and momenta subject to the corresponding  Hamiltonian constraint. For these cases we specify the set $\{\nu(0),\, p_\nu(0), \beta_\pm(0),\,p_-(0)\}$, and also $\{\phi(0),p_\phi(0)\}$ for non-zero scalar field, with $p_+(0)$ determined by the Hamiltonian constraint. This method provides identical data for all variables except for $p_+(0)$, since the Hamiltonian constraints differ. 

In contrast, the initial data for the quantum scalar field cases $H_Q$ and $H_Q^{\rm{eff}}$ require the initial quantum state  of the scalar field; the data is $\{\nu(0),\, p_\nu(0), \beta_\pm(0),\,p_-(0), |\psi\rangle(0)\}$, again with $p_+(0)$ determined by solving the respective classical and effective hamiltonian constraints. For comparison with the classical scalar field cases, we fix the data for the latter by setting
\bea
\phi(0) =\sqrt{\langle \hat{\phi}^2\rangle(0)}; \quad
p_\phi(0) =\sqrt{\langle \hat{p}^2_\phi\rangle(0)},
\label{cdata}
\eea
with the expectation values taken in the initial state.  Lastly, for all illustrative numerical evolutions, we set the scalar field's mass to $m=0.01$ and polymer scale $\lambda = 0.1$. For the quantum scalar field we work in the fixed oscillator basis $\{|n\rangle_0\}$ at the initial time so that the evolved state at a subsequent time $t$ is 
\bea 
\ket{\psi}(t)=\sum_n c_n(t)\ket{n}_{t=0}.
\label{q-state}
\eea 
For state evolution, we write the scalar field Hamiltonian operator as a matrix in the oscillator representation with a truncation at level $n=30$; the Schr\"{o}dinger equation then becomes a set of $30$ coupled equations for the coefficients $c_n(t)$.  The initial state is chosen to be the ground state at the initial time. We confirm that changing truncation level does not significantly affect our results provided the initial state is the ground state, and the evolution time is such that probability remains conserved. 

The following figures provide a sample of our numerical results. We have verified that the respective Hamiltonian constraints remain conserved up to order $10^{-9}$ and that the probability  is conserved to a similar order for the quantum scalar field case.  

\subsection{Volume, anisotropy and Kasner sum evolution}

Fig.\ref{fig01:dynamics} shows the evolution of volume $\nu$ (first row), anisotropies (second row), and the log of the sum of the squared Kasner exponents $\sum_k p_k^2$ (third row), for the vacuum, and for the classical and quantum scalar fields coupled to (classical and effective) gravity; the initial state for the latter is the vacuum $|0\rangle_{t=0}$.  For each sub-figure, comparable initial data is chosen as described above; for all cases the data is $\{\nu(0)=10, \, \beta_+(0)= -0.05,\,\beta_-(0)= 0.01,\, p_-(0)=-10,\,p_\nu(0)=-5\}$, with $p_+(0)$ determined by the respective hamiltonian constraints; the simulations run for the same time (with lapse $N=1$); the dashed blue line is the case where the gravitational hamiltonian is the classical (not the effective) one, which terminates in a singularity, e.g. near $t=40$ in the first two frames, and near $t=55$ in the third frame of the first row. The small differences in the location of the bounce and singularity in these figures is due to the difference in the initial value $p_+(0)$ as determined by the respective Hamiltonian constraints.

\underbar{{\it Volume bounces:}} The first row shows that the bounces are asymmetric for each of the effective cases (vacuum, classical and quantum scalar field), with the most bounces occurring for the vacuum case; each case also indicates a higher rise in volume after a series of relatively smaller bounces and expansions; the quantum scalar case differs significantly from the classical scalar one in the number of initial bounces.  

\underbar{{\it Anisotropies:}} The second row compares the anisotropies $\beta_\pm$ in each of the cases. The approach to the first bounce is noteworthy: all cases are very similar as the first bounce is approached, with anisotropies vanishing near $t=40$. However after the bounce, the second and third frames (the cases $H^{\rm{eff}}_C$ and $H^{\rm{eff}}_Q$) show higher anisotropies that rise rapidly near each bounce; it is also evident that the classical and quantum scalar field cases differ significantly in this respect. Fig. \ref{fig09:zoomed_in_anisotropies} is a view of the anisotropies at the first bounce point of Fig. \ref{fig01:dynamics}; it shows that all three bounce cases are qualitatively similar. The evolutions of $\beta_+$ and $\beta_-$ are quite different due to their definitions in (\ref{bp}) and (\ref{bm}): the former is a function of all three scale factors, while the latter depends on only two. (The different evolutions are a feature also of the exact Kasner solution where $\beta_+ \sim \ln(t^{p_1+p_2-p_3^2})$ and $\beta_+ \sim \ln(t^{p_1-p_2})$, for Kasner exponents $p_1,p_2,p_3$ satisfying the sum rule.)  

\underbar{{\it Kasner sum rule:}} The third row of Fig.\ref{fig01:dynamics} shows zoomed in plots of log of the Kasner sum  $\sum_{i=1}^3 (p_i^2)$ and $\sum_{i=1}^3 (p_i^{\rm{eff}\,2})$(blue and orange curves respectively), and log of the effective volume $\nu^{\rm eff}$ (green curve), as the singularity/bounce point is approached. The blue curve shows that the Kasner sum tends to unity as the classical singularity is approached for each of the vacuum, classical, and quantum scalar field cases ($H_{IX},H_C,$ and $H_Q$ ), supporting the BKL conjecture that ``matter does not matter" as the singularity is approached; (the dip in these curves very close to the singularity point is due to break down of numerical evolution very close to the singularity). 

However, the orange curves (corresponding to $H_{IX}^{\rm{eff}},\ H_C^{\rm{eff}}$ and  $H_Q^{\rm{eff}}$ ---where there is a bounce) show that the sum rule is a constant away from the bounce point, but not unity due to the polymer scale $\lambda=0.1$; it undergoes oscillations through the bounce (where the green volume curve dips); the differences in the three frames in this row show that ``matter matters" when the universe bounces, and in particular whether matter is classical or quantum.  Furthermore, the oscillations in the sum rule indicate that the approach to the volume bounce is not Kasner. This is in contrast with the BKL conjecture, which states that the approach to the classical singularity is Kasner-like; the conjecture is of course silent on the singularity free bouncing cosmologies we study here.

\subsection{Scalar field excitation and density}

Fig. \ref{fig02:excitations} gives the results for the evolution of the scalar field quantum state, the coefficients $|c_n(t)|^2$ in (\ref{q-state}) for the cases $H_Q$ (classical gravity with quantum scalar field) in the first row, and $H_{Q}^{{\rm eff}}$ (effective gravity with quantum scaler field) in the second row, with initial state $ |0\rangle_{t=0}$; the first and second columns show the even and odd level excitations (for Hilbert space truncation $n=30$ in the fixed basis chosen at $t=0$). 

For $H_Q$ (first row), the initial data is the same as in the for the dashed blue curve in the rightmost column of Fig. \ref{fig01:dynamics}; the singularity is just before $t=60$. The left frame shows the even state $|2k\rangle$ excitations (which connect with the initial state $|0\rangle_{t=0}$ due to the quadratic Hamiltonian); as the singularity is approached, it is evident that higher states $k=2,4,6$ are excited to nearly equal levels $|c_k|^2 \in (0.1,0.2)$, but the states near the truncation point $n=30$ are not significantly excited; the right frame shows the non-excitation of the odd states $|2k+1\rangle$ to order $10^{-14}$. The total probability is numerically conserved to a similar order. 

For $H_Q^{\rm{eff}}$ (second row), the initial data is the same as in the rightmost column of Fig. \ref{fig01:dynamics} (with the four bounces).  The left frame shows the even excitations; the higher levels are excited through each bounce; $|c_0|^2$ drops and $|c_k|^2 \ (k=2,4,6)$ rise; at later times these higher levels remain excited and the scalar field never returns to purely the ground state at maximum expansion. The right frame again shows that the odd levels are not excited to order $10^{-10}$, and again total probability is numerically conserved.  

Fig. \ref{fig03:exciations_detail} shows a zoomed in version of the even state excitations of the scalar field for $H_q^{\rm{eff}}$ near the first bounce; the left frame shows the lower levels, and the right frame shows the higher levels near the truncation $n=30$; the log of the volume (dashed red line) is overlaid. This shows order $10^{-2}$ excitations of the higher levels with a similar pattern.


These results may be viewed as the quantum mechanics analog of particle production on a fixed time dependent background, with the difference that our calculation includes non-perturbative back reaction of the evolving scalar field state on the evolution of the gravitational variables.  

\begin{figure*}[hbt]
    \centering
   \includegraphics[width=0.85\textwidth]{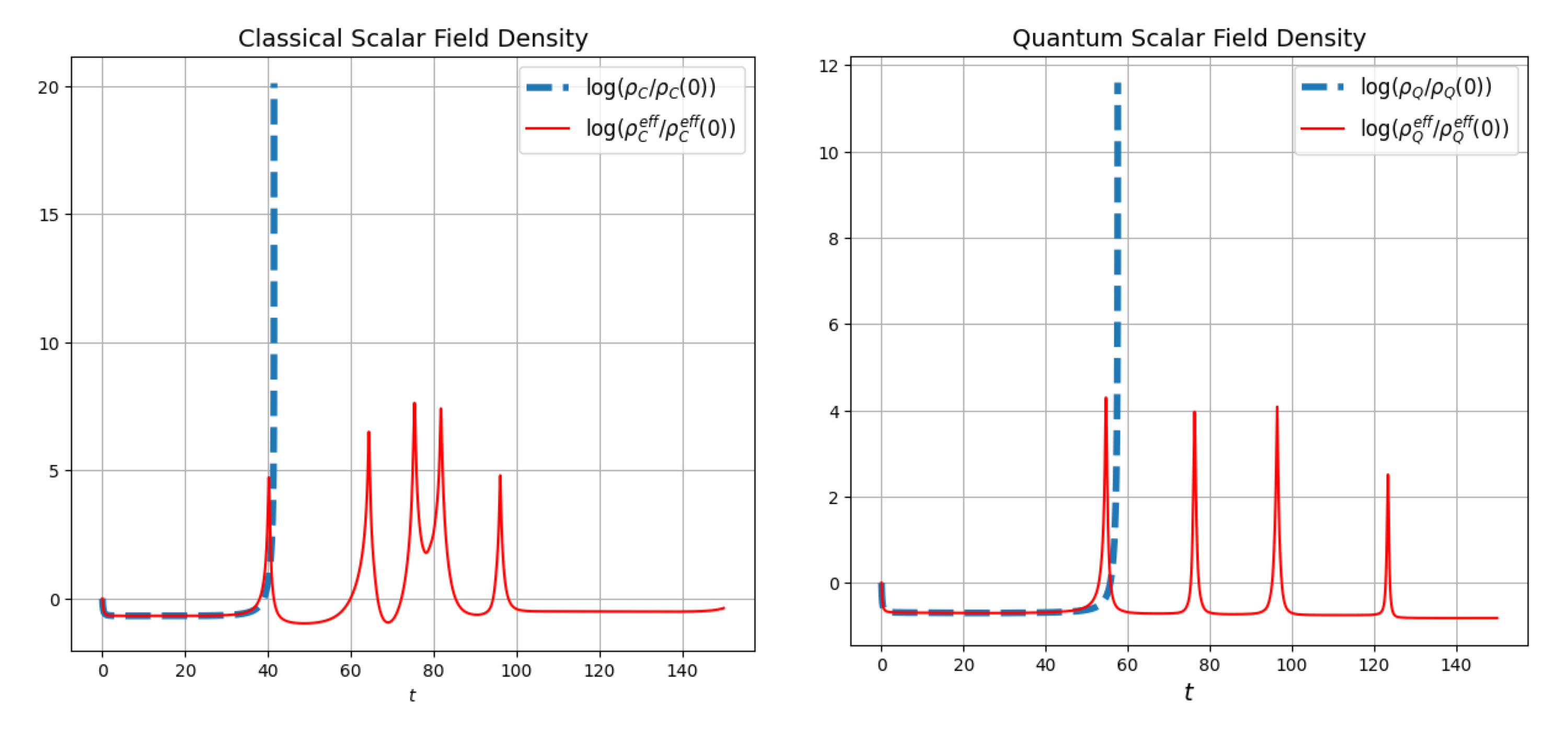}
    \caption{Scalar field densities in a classical  and effective universe. In the former case (dashed blue line) the density diverges at the singularity; in the latter cases (orange line) the density spikes at a maximum at each bounce shown in Fig. \ref{fig01:dynamics}  }
    \label{fig03:densities}
\end{figure*}

 Fig. \ref{fig03:densities} shows the evolution of the classical and quantum scalar field densities for $H_C^{\rm{eff}}$ and $H_Q^{\rm{eff}}$, 
 \bea 
\rho &=& \frac{1}{\nu} H_\phi =\frac{p_\phi^2}{2\nu^2}+\frac{1}{2}m^2\phi^2,\\
\rho_Q &=& \frac{1}{\nu}  \langle \hat H_\phi \rangle =\frac{\langle \hat p_\phi^2 \rangle}{2\nu^2}+\frac{1}{2}m^2\langle \hat \phi^2\rangle,
\eea 
again for the same initial data as in Fig. \ref{fig01:dynamics}, with the cases $H_C$ and $H_Q$ (no bounce classical gravity) overlaid (dashed blue line). The bounded spikes in the densities are evident at each of the bounce points.   

\begin{figure*}[t]
    \centering
   \includegraphics[width=1.0\linewidth]{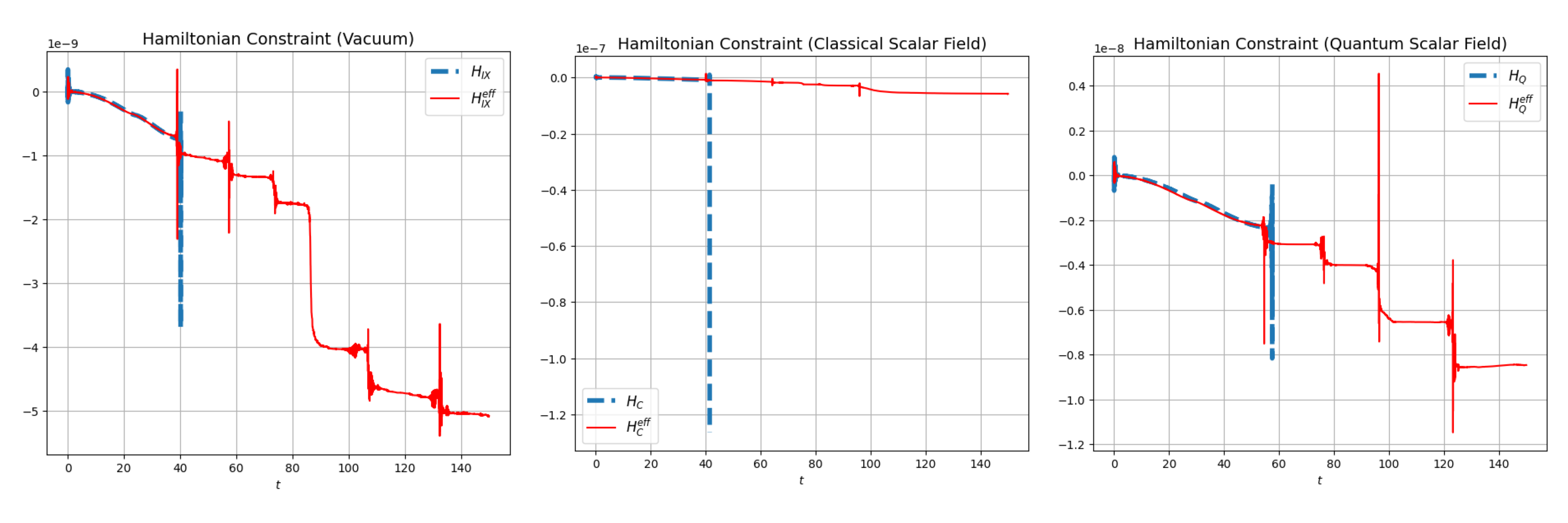}
    \caption{Numerical conservation of the Hamiltonian constraint for the evolutions shown in Figs. \ref{fig01:dynamics}, \ref{fig02:excitations}, and \ref{fig03:densities}. }
    \label{fig04:ham_cons}
\end{figure*}

Lastly, Fig. \ref{fig04:ham_cons} exhibits the conservation of the Hamiltonian constraint for each of the cases we study to order at least to order $10^{-7}$. This shows the accuracy of our numerical evolutions, including the oscillations of the Kasner sum at the bounce point in Fig. \ref{fig01:dynamics}. It also reflects probability conservation of the quantum scalar field case (third frame), where any probability violation would result in non-conservation of the Hamiltonian constraint through the expectation value of the matter contribution to this constraint.  
 
\subsection{Chaos}
Close to the singularity, the Mixmaster Universe is known to have chaotic dynamics. We study this comparatively for the six models using two methods:  computing Lyupanov exponents and the fractal dimension of trajectories in phase space.  

\subsubsection{Lyapunov exponents}

One measure of chaos is the Lyapunov exponent: given two infinitesimally close initial data points $x_1(0)$ and  $x_2(0)$ for a dynamical system, if the initial separation $\xi(0)=|x_2(0)-x_1(0)|$ grows as $\xi(t) \sim \xi(0)e^{\lambda t}$ with 
$\lambda > 0$ the system is considered to be chaotic, and $\lambda$ is called the Lyapunov exponent. An $n-$dimensional phase space has $n$ such exponents. One way to compute them is to linearize the dynamical system and solve the system
\bea 
\frac{d}{dt}\xi_n(t)=J(f(x(t)))_{nm}\xi_m(t); \label{perturbevol}
\eea
where $J$ is the matrix that arises from linearization of the evolution equations. The Lyapunov exponents are  given by  
\bea 
\lambda_n =\lim_{t \to \infty}\ln\bigg[\frac{\xi_n(t)}{\xi_n(0)}\bigg]
\eea 

In the numeric integration of (\ref{perturbevol}), it often happens that the $\xi_n(t)$ diverge in a short duration. To remedy this, Bennetin et al \cite{benettin1976kolmogorov} introduced the the idea of re-orthonormalizing the  vectors $\xi_n(t)$ at regular time intervals, and storing their corresponding scalings. This is the method we use to calculate the Lyapunov exponents for the six models studied.   The results appear in Table \ref{Lexp}. 
\begin{table}
\begin{tabular}{|c|c|}
\hline
\textbf{System} & \textbf{Lyapunov Exponent} \\
\hline
$H_{IX}$ & $3.210$ \\
\hline
$H_{IX}^{\rm{eff}}$ & $0.4093$\\
\hline
$H_C$ & $4.4320$ \\
\hline
$H_C^{\rm{eff}}$ & $2.100$\\
\hline
$H_Q$ & $4.4320$ \\
\hline
$H_Q^{\rm{eff}}$ & $1.024$\\
\hline
\end{tabular}
\caption{Maximal Lyapunov Exponents for the six models}
\label{Lexp}
\end{table}

\begin{figure*}[hbtp]
    \centering
   \includegraphics[width=1\textwidth]{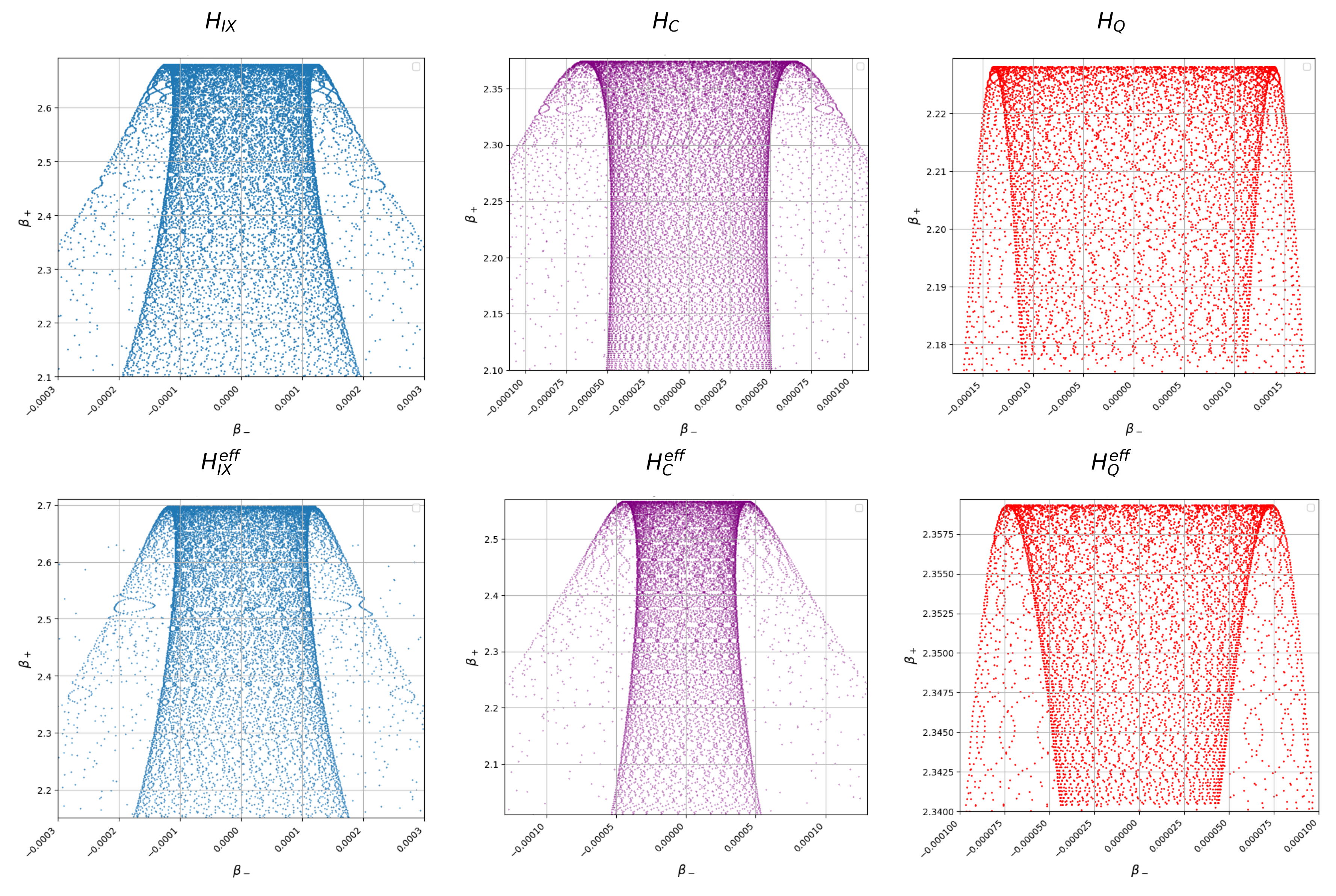}
    \caption{Zoomed-in phase plots of the anisotropies $\beta_\pm$ for each of the six models studied. Self-similarity and fractal-like structure is evident in each case; all  are qualitatively similar, but with notable differences for the quantum scalar field.}
    \label{fig04:phaseplots}
\end{figure*}

These are computed from runs up to  $t=200$, keeping in mind that  infinitesimally close initial trajectories first diverge, and then settle into a constant distance on average, forming an attractor set. The dimensions of the matrix are different in each of the six cases due to the number of dynamical equations, with the largest dimension occurring for the quantum scalar field cases with truncation (at $n=30$). Notable in the table is that the Lyapunov exponents for the effective gravity cases are smaller than for classical gravity, providing evidence that chaos is reduced, at least by this measure. In particular there is a nearly factor of four difference in the exponents between $H_C$ (classical gravity with classical scalar field) and $H_Q^{\rm{eff}}$  (effective gravity with quantum scalar field).

 \begin{figure*}[!htbp]
    \centering
    \includegraphics[width=1.0\linewidth]{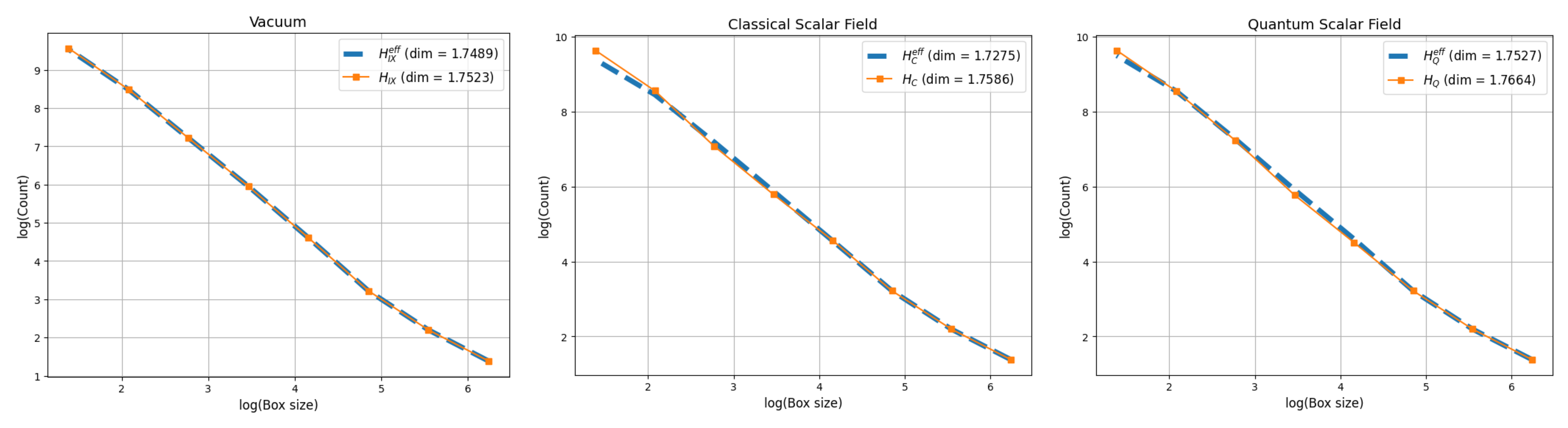}
    \caption{We see here the results of the box-counting analysis to compute the dimensions of the plots in Figure \ref{fig04:phaseplots}. For each plot in the figure, we use the polymerized and unpolymerized counterparts for each species (vacuum, classical scalar field, and quantum scalar field). We see that in each case, polymerization reduces the fractal dimension of the image, rendering the system to be less chaotic.}
    \label{fig:fractal_dims}
\end{figure*}

\subsubsection{Fractal dimension}

There are several methods for estimating the fractal dimension of attractor sets in phase space \cite{strogatz2018nonlinear,CornishLevin}. One of these is the so-called box counting method computed from images of the fractal. For a fractal set $F$, the number $N$ of boxes of size $\epsilon$ it takes to cover $F$ scales as
 \bea 
\log(N(\epsilon)) = -D \log(\epsilon).\label{box_count_dim}
\eea  
We apply this to the phase space plots  in Fig. \ref{fig04:phaseplots},  which show evolution in the $\beta_+-\beta_-$ plane for each of the six Hamiltonians we studied. Specifically, we place a sequence of finer and finer square grids on the gray scale images in this figure, and count the number of pixels above a certain gray scale threshold ($90\%$) in each grid box. The aim is to obtain a comparison of fractal dimension for each of the six cases to see how they compare with the Lyapunov exponents, for a relative ranking of degree of chaos.

The results of this are shown in Fig. \ref{fig:fractal_dims}. The fractal dimensions are all very close to one another in the range $[1.72, 1.77]$, but with the effective Hamiltonian cases having slightly smaller dimensions than the classical case. As such this method does not provide a significant differentiation of chaos in the six models we studied.

Our main result of this subsection is evidence of significant chaos reduction in the cases where there is a bounce, as exhibited in Table \ref{Lexp}. These results are  intuitively reasonable from the perspective that a minimum volume effectively saturates the number of collisions of the universe particle with the walls of the Bianchi IX potential, compared to the case where there is no minimal volume and the singularity is present. This in turn means that a more pervasive exploration of the phase space is thwarted by the bounce in volume, thereby reducing the degree of chaos. As this intuition is broad, it likely applies to {\t any} mechanism that replaces the singularity with a bounce; see e.g. \cite{Moriconi:2014wea,Barca:2024goe} for other approaches where chaos is either reduced or eliminated altogether. 
 
\section{Summary and Discussion}

We presented a comparative study of six Bianchi-IX models, with classical and effective gravity, and with classical and quantum massive scalar field. The effective gravity case used space discreteness only in the volume variable, in a manner similar to that used in LQC models \cite{blackmore2023numerics}. We saw that this is sufficient for producing bouncing dynamics, both analytically and numerically. The model with the quantum scalar field used the Schr\"{o}dinger equation for state evolution, and a modified Hamiltonian constraint where the scalar field contribution is replaced by its expectation value; this is a non-perturbative and self-consistent method for back reaction. 

Our simulations show a number interesting features: (i) multiple bounces occur with the effective gravity hamiltonian cases, with expansion to larger volume after successive bounces; (ii) all bounces are asymmetric; (iii) anisotropies are relatively larger with classical and quantum matter than in vacuum as the universe emerges from a bounce; (iv) away from the bounce, the effective gravity quadratic Kasner exponent sum is a constant of the motion, but undergoes oscillations through the bounce, whereas it is conserved and approaches unity (with or without matter) in the classical case; (v) the quantum scalar field undergoes excitations from the initial ground state as the singularity/bounce is approached, and partial de-excitation after the bounce; (vi) chaos persists in the effective gravity case, but the  Lyapunov exponents indicate that it is ``reduced" in the effective gravity cases,  with the largest reduction occurring for the vacuum effective case $H_{IX}^{\rm{eff}}$, a result at least qualitatively consistent with results for vacuum Bianchi IX using entirely different methods \cite{brizuelabojowald}; (vii) the box counting method in phase space also provides evidence for chaos with a fractal dimension of between 1.72 and 1.77 for all six models, but with only marginal differences between them.  

As noted above, the observation that chaos is reduced in effective gravity is not surprising, since chaos is produced by the increasing number of bounces off the triangular potential walls as the singularity is approached. If the universe achieves a minimum volume, the potential walls stop closing in, and it is reasonable to expect that the number of collisions with the walls reaches a maximum, and then starts to reduce through the bounce. The presence of matter affects this only in the transfer of some ``energy" between the matter and gravity terms in the Hamiltonian constraint; the reduction in chaos appears to be mainly due to the bounce in the effective gravity cases. This argument also suggests that chaos is not merely a coordinate artefact, and that although Lyapunov exponents may differ if other phase space variables and time coordinates are used, it appears unlikely that all exponents would become zero or negative, given the nature of the potential.

This work extends the much-studied vacuum Bianchi IX model in two ways -- by introducing effective gravity via fundamental discreteness corrections and quantum matter with back reaction. The same methods may be used for the other Bianchi models, with their somewhat simpler potentials. More interesting from a quantum gravity perspective are inhomogeneous models, the simplest of which are the Gowdy cosmologies with one local degree of freedom. Introducing quantum matter with back reaction in the manner we have done here would require introducing matter expectation value terms in both the  hamiltonian and diffeomorphism constraints, and ensuring their algebra remains consistent. This remains to be seen.  

\smallskip
\noindent \underbar{Acknowledgements} This work was supported by the Natural Science and Engineering Research Council of Canada.

\bibliography{references.bib}  

\end{document}